\documentclass[aps, twocolumn,noshowpacs,preprintnumbers,amsmath,amssymb]{revtex4}

\usepackage{verbatim}
\usepackage{wasysym}

\usepackage{graphicx}
\usepackage{dcolumn}
\usepackage{bm}
\usepackage{wrapfig}

\begin{document}

\title{Morphology dependence of radial elasticity in multiwalled BN nanotubes}

\author{Hsiang-Chih Chiu}
\author{Suenne Kim}
\affiliation{School of physics, Georgia Institute of Technology, Atlanta, USA}
\author{Christian Klinke}
\affiliation{Institute of Physical Chemistry, University of Hamburg, Hamburg, Germany}
\author{Elisa Riedo}
\affiliation{School of physics, Georgia Institute of Technology, Atlanta, USA }

\begin{abstract} 

We report on the measurement of the effective radial modulus of multiwalled Boron Nitride nanotubes with external radii in the range 3.7 to 36 nm and number of layers in between 5 and 48. These Boron Nitride nanotubes are radially much stiffer than previously reported thinner and smaller Boron Nitride nanotubes. Here, we show the key role of the morphology of the nanotubes in determining their radial rigidity, in particular we find that the external and internal radii, $R_{ext}$ and $R_{int}$, have a stronger influence on the radial modulus than the NT's thickness, $t$. We find that the effective radial modulus decreases nonlinearly with $1/R_{ext}$ until reaching, for a large number of layers and a large radius, the transverse elastic modulus of bulk hexagonal-Boron Nitride. 

\end{abstract}

\maketitle

Boron nitride nanotubes (BN-NTs) are structurally similar to Carbon nanotubes (C-NTs) and can be considered as rolled up hexagonal sheets of alternating boron and nitride atoms in a honeycomb structure. BN-NTs are electronically insulating with a large bandgap of ca. 5.5 eV, independently of their radius, layers and chirality \cite{1}. They are thermally highly conductive, chemically inert and more resistant to oxidation compared to C-NTs \cite{2,3}. In addition, they are theoretically predicted to possess piezoelectricity \cite{4,5}. These exceptional properties enable BN nanotubes and nanosheets to find potential applications in various fields such as reinforcement in composite materials, opto-electronic nanodevices, and hydrogen storages. Their electrical insulating properties also make them ideal substrates for C-NT and graphene due to perfect lattice matching \cite{6}. A comprehensive review about BN nanomaterials can be found in Ref. [7].  Theoretical studies have shown that the axial Young's modulus of multiwalled (MW) BN-NTs is about 0.7 times that one of C-NT \cite{8,9}. It is generally expected that the Young's modulus of BN-NTs is smaller than that one of C-NTs for the same radius, and the same number of layers, since the covalent C-C bonds are stronger than the covalent and polar B-N bonds. The axial Young's modulus of a MW BN-NT with a radius of 3.5 nm has been measured by means of the thermal excitation method and values around 1.2 TPa have been found \cite{10}. Chemical vapor deposited (CVD) MW BN-NTs with external radii, $R_{ext}$, ranging between 10 and 93 nm display axial Young's moduli in the range 0.5-1 TPa \cite{11,12,13,14,15}.

\begin{figure}[ht]
  \centering
  \includegraphics[width=0.4\textwidth]{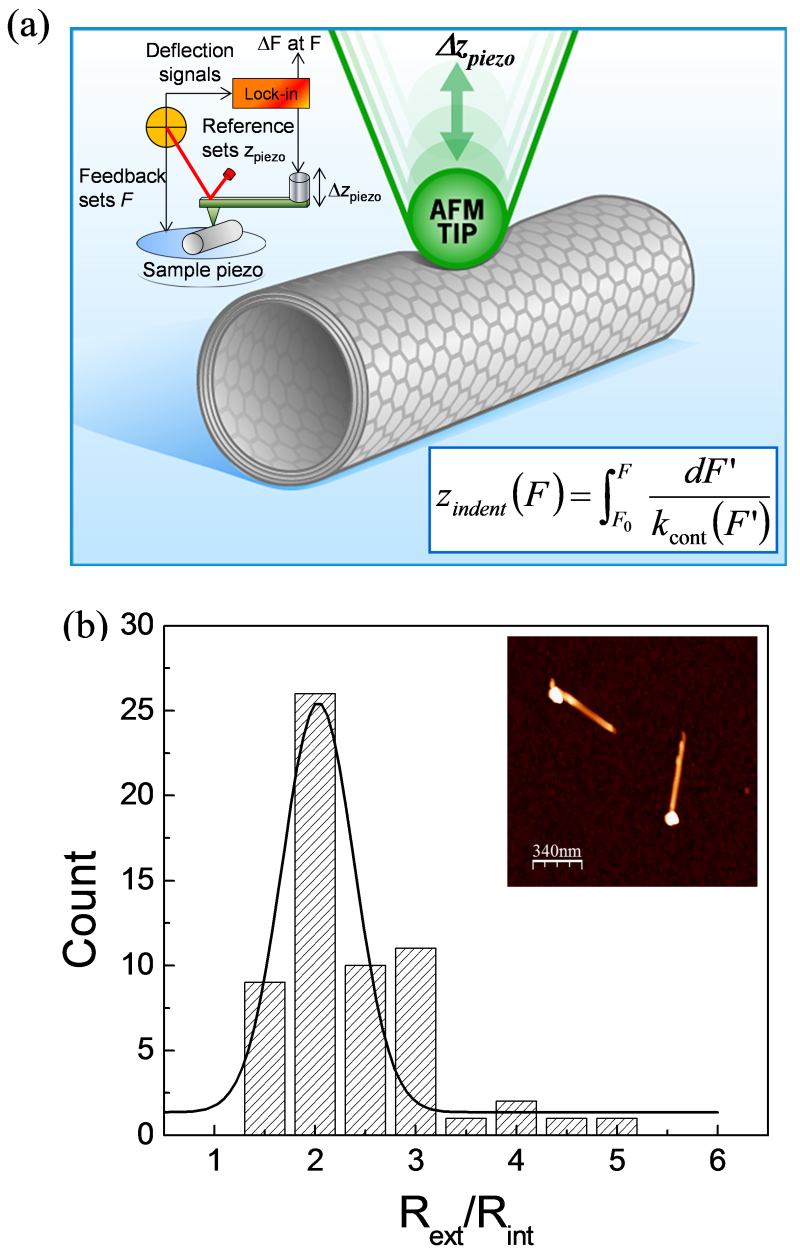}
  \caption{\textit{(a) A cartoon shows the experiment setup for the MoNI measurement; (b) The statistics of $R_{ext}$/$R_{int}$ as obtained by TEM from 33 MW BN-NTs. The center from the Gaussian fit is 2.0 $\pm$ 0.3. The insect shows a typical AFM image of BN-NTs deposited on a silicon substrate.}}
\end{figure}

The radial elasticity and deformation mechanisms of BN- and C-NTs are important but not fully explored and understood. For example, the radial stiffness of C-NTs measured so far varies over three orders of magnitude without a clear understanding of the reason \cite{16,17}. Nevertheless the radial deformation of nanotubes strongly affects their physical and structural properties. For CVD C-NTs, it has been shown that the radial deformations have significant impact on their electrical properties \cite{18}. A recent study also demonstrates that insulating BN-NTs can be transformed into semiconductors when the nanotube is deformed \cite{19}. The radial deformation of nanotubes is a complex subject, because differently from the axial Young modulus, the effective radial modulus is not easily defined \cite{20}. The nanotubes can deform under localized radial compression in different ways depending on the degree of compression and nanotube morphology, e.g. number of layers, radius, length, and capped ends. Furthermore BN-NTs and C-NTs are not isotropic, i.e. the in plane and out of plane elastic constants are very different in both BN and graphite. Finally, different elastic deformation mechanisms can occur when the tube is subject to radial localized compression \cite{20,21}. For example, for large compressing forces, and thin tubes, deformation and ovalization of the nanotube's cross section will take place; whereas for small indentations, and thick tubes, such compression is undertaken mainly by the walls of nanotube. A critical parameter controlling the deformation's mechanism of a cylindrical shell under radial pressure is usually the tube's thickness to external radius ratio, $t/R_{ext}$, whose importance has been indicated in classical shell theory \cite{22}. Recent atomic force microscopy (AFM) based experiments on single walled (SW) and MW BN-NTs have reported values of radial elasticity in the range 2 to 14 GPa for MW BN-NTs with radii decreasing from 3.3 nm to 0.3 nm respectively, and values up to 40 GPa for SW BN-NT with radii below 1 nm \cite{17,23}. Another AFM study has shown that, for a MW C-NT with a radius of 4.5 nm and 6 layers, during the load-displacement measurement, the C-NT undergoes initially ovalization and then is totally flattened for loads larger than 40 nN, and the effective radial modulus varies depending on the pressure regime during the compression process \cite{21}. 

Here, we use the AFM based modulated nano-indentation (MoNI) technique \cite{16,24} (Fig. 1(a)) to investigate the radial stiffness of individual CVD grown MW BN-NTs with radii ranging between 3.7 and 36 nm, and with a wall thickness approximately equal to 0.5 times the external radius, $R_{ext}$, as obtained by the statistical TEM analysis shown in Fig. 1(b). The radial deformation is here always measured during the unloading regime for indentations below 10\% of $R_{ext}$ to insure the probing of the local radial elasticity of the nanotubes. The effective radial moduli, $E_{Radial}$, are obtained from the force versus indentation curves using the Hertz contact mechanics model, modified to include adhesion forces. We find that $E_{Radial}$ of MW BN-NTs decreases with increasing $R_{ext}$ and the number of layers, $L$, from about 192.2 $\pm$ 46.8 GPa for $R_{ext} = $ 3.7 $\pm$ 0.1 nm and $L = $ 5.4 $\pm$ 1, to a plateau value of 24 $\pm$ 11 GPa. These elastic moduli are larger than previously reported values for MW BN-NTs \cite{23,25} because of the small indentation and large number of wall layers in the BN-NTs studied in this letter. When probing thick tubes the ovalization process at small indentations as in our experiments is minimal, instead, the localized radial compression creates a dimple in the tube's top layers and the elasticity depends on the layer-layer interaction and layer curvature. Here, we demonstrate the key role of the morphology in determining the radial rigidity of MW BN-NTs, in particular we find that the external and internal radii, $R_{ext}$ and $R_{int}$, have a stronger influence on the radial modulus than the thickness of the nanotubes, $t$. We propose that the radial modulus varies with the morphology of the nanotubes as $E_{Radial} = A / (R_{ext}^2 (R_{int} - 0.34^2)) + E_{\circ}$, where $A$ and $E_{\circ}$ are two constants. For the regime of morphology investigated here, we find that $E_{\circ} = $ 20.3 $\pm$ 3.4 GPa for the BN-NTs and $E_{\circ} = $ 27.9 $\pm$ 6.0 GPa for C-NTs with similar morphology. These values are very close to the elastic constants along the $C_{33}$ axis of the corresponding bulk materials, i.e., 27 GPa for hexagonal Boron Nitride ($h$-BN) \cite{26} and 36 GPa for Highly ordered pyrolytic graphite (HOPG) \cite{27}, while the elastic modulus along the direction of $C_{33}$ of HOPG, namely $E_{33}$, is reported to be about 28-31 GPa \cite{28}.

Both AFM based standard indentation curves \cite{29} and the MoNI technique (Fig. 1 (a)) are performed on individual MW BN-NT deposited on a clean Silicon substrate to obtain their radial elasticity. The AFM used here is a Veeco Nanoscope 3a Multimode system. For standard force $vs.$ indentation curves we use soft AFM cantilevers with a normal spring constant of about $k_{lev} \approx $ 0.2 N/m and tip radius $R_{tip} \approx $ 50 nm. For the MoNI experiment, we use a silicon AFM tip (Nano and More, ppp-NCHR) with a typical tip radius $R_{tip} \approx $ 40 nm, attached to a stiff cantilever with spring constant $k_{lev} \approx $ 40 N/m, calibrated with the Sadar's method \cite{30}. The CVD grown MW BN-NTs are purchased from NanoTechLabs, Inc. (Yadkinville, NC). They are suspended in IPA (Isopropyl alcohol) within an ultrasonic bath. A drop of this solution is placed on a Silicon wafer for 30 seconds and blown dried with compressed nitrogen gas. The BN-NTs remain anchored to the Silicon surface via van der Waals forces. This procedure is repeated several times until the appropriate density of nanotubes on the surface is reached for AFM measurements. The dimensions of the BN-NTs are characterized by both TEM and AFM. The TEM images of BN-NT and CVD C-NT used in Ref. 15 are reported in Ref. 28 for comparison. In addition, as determined by TEM, the BN-NT used here have an average external to internal radii $R_{ext}$/$R_{int} = $ 2.0 $\pm$ 0.48, determined from the Gaussian fit to the data shown in Fig. 1 (b). In order to represent more accurately the impact of such data distribution, for $R_{ext}$/$R_{int}$ we choose the Full Width Half Maximum (FWHM), 0.48, of the Gaussian distribution as its error bar. This almost constant $R_{ext}$/$R_{int}$ ratio permits us to compare the BN-NT results to those obtained for C-NT in Ref. [16], with constant $R_{ext}$/$R_{int}$ ratio equal to 2.2 $\pm$ 0.2. The values of $R_{ext}$ and $R_{tip}$ are directly inferred from the topographical images of the BN-NT by using the equation $R_{tip} = w^2/(16 \cdot R_{ext})$, where $w$ is the apparent width of the AFM imaged nanotubes.

During a typical MoNI experiment, an AFM tip, which is vertically oscillated at a fixed frequency with sub-nanometer amplitudes, applies a localized radial pressure on BN-NT lying on a Si substrate. The oscillations are applied to the AFM tip via a piezoelectric stage rigidly attached to the cantilever, and controlled by a Lock-in amplifier (Stanford Research Systems, SR830), while a constant normal force $F$ between the tip and the NT is maintained by the feedback loop of the AFM (see Fig. 1 (a)). In order to remain in the linear elastic regime, the piezo-stage oscillations are chosen to be only 1.0 \AA. The oscillation frequency is chosen to be 0.7523 kHz with optimized feedback parameters. 

During the indentation, the fixed piezo-stage oscillation amplitude $\Delta z _{piezo}$ is equal to the sum of the cantilever bending and tip-nanotube normal deformation. Under such circumstances, the AFM cantilever and the tip-nanotube contact can be considered as two springs connected in series: the cantilever with stiffness $k_{lev}$ and the tip-nanotube contact with stiffness $k_{cont}$. The force required to stretch these two springs in series with a total displacement $\Delta z _{piezo}$ is equal to the normal force variation $\Delta F$. This experimental configuration allows us to measure the total stiffness $k_{tot}$ at each normal load $F$, fixed by the feedback loop of the AFM, from the following relation:

\begin{equation}
\frac{\Delta F}{\Delta z _{piezo}} = k_{tot} (F) = \left( \frac{1}{k_{lev}} + \frac{1}{k_{cont}} \right)^{-1}
\end{equation}

Since $k_{lev}$ is known, the measurement of $\Delta F$/$\Delta z _{piezo}$ at different normal loads $F$ allow us to acquire the radial stiffness $k_{cont}$ as a function of $F$. Figure 2 (a) shows a typical $k_{cont}(F)$ obtained from a BN-NT with $R_{ext} = $ 13.7 $\pm$ 0.4 nm. The data are acquired automatically through a home written program at each constant normal load $F$. The negative values of $F$ indicate the presence of the adhesion force $F_{Adh}$. Before obtaining the radial Young modulus of the nanotubes, we extract force $vs.$ indentation curves by integrating the equation $\Delta F$/$k_{cont}(F) = \Delta z_{indent}$. Figure 2 (b) shows the results for 4 different MW BN-NTs with $R_{ext} = $ 3.7 $\pm$ 0.1 nm, 5.1 $\pm$ 0.1 nm, 6.0 $\pm$ 0.1 nm and 16.9 $\pm$ 0.2 nm. The radial stiffness defined as $d F$/$d z_{indent}$ is clearly increasing with decreasing $R_{ext}$ within the indentation range. The 3/2 power law dependence between $F$ and $z_{indent}$ as predicted by the Hertz model is also clearly satisfied, indicating that the main deformation mechanism here is not the ovalization of the NT which would give a linear dependence between $F$ $vs.$ $z_{indent}$, as reported in Ref. [21, 25]. We note that MoNI, differently from standard static compression AFM experiments, permits to perform elasticity measurements with indentations as small as fractions of an Angstrom. This is particularly important for small NT to ensure that it is not undergoing damage or plastic deformations. For comparison, in Ref. [29] we add standard force $vs.$ indentation ($F$ $vs.$ $z$) curves during the loading and unloading process, using a very soft cantilever to insure small indentations. During the indentation the load is kept below 10 nN, thus the radial compression is mainly performed by the walls of the nanotube, with negligible or minimum radial ovalization. These $F$ $vs.$ $z$ curves show that the investigated BN-NT with $R_{ext} \approx $ 25 nm has a very high stiffness, but it is not possible to obtain an accurate measure of the contact stiffness, and hence of the radial modulus, with such soft cantilevers and standard force $vs.$ indentation measurements.

\begin{figure}[ht]
  \centering
  \includegraphics[width=0.4\textwidth]{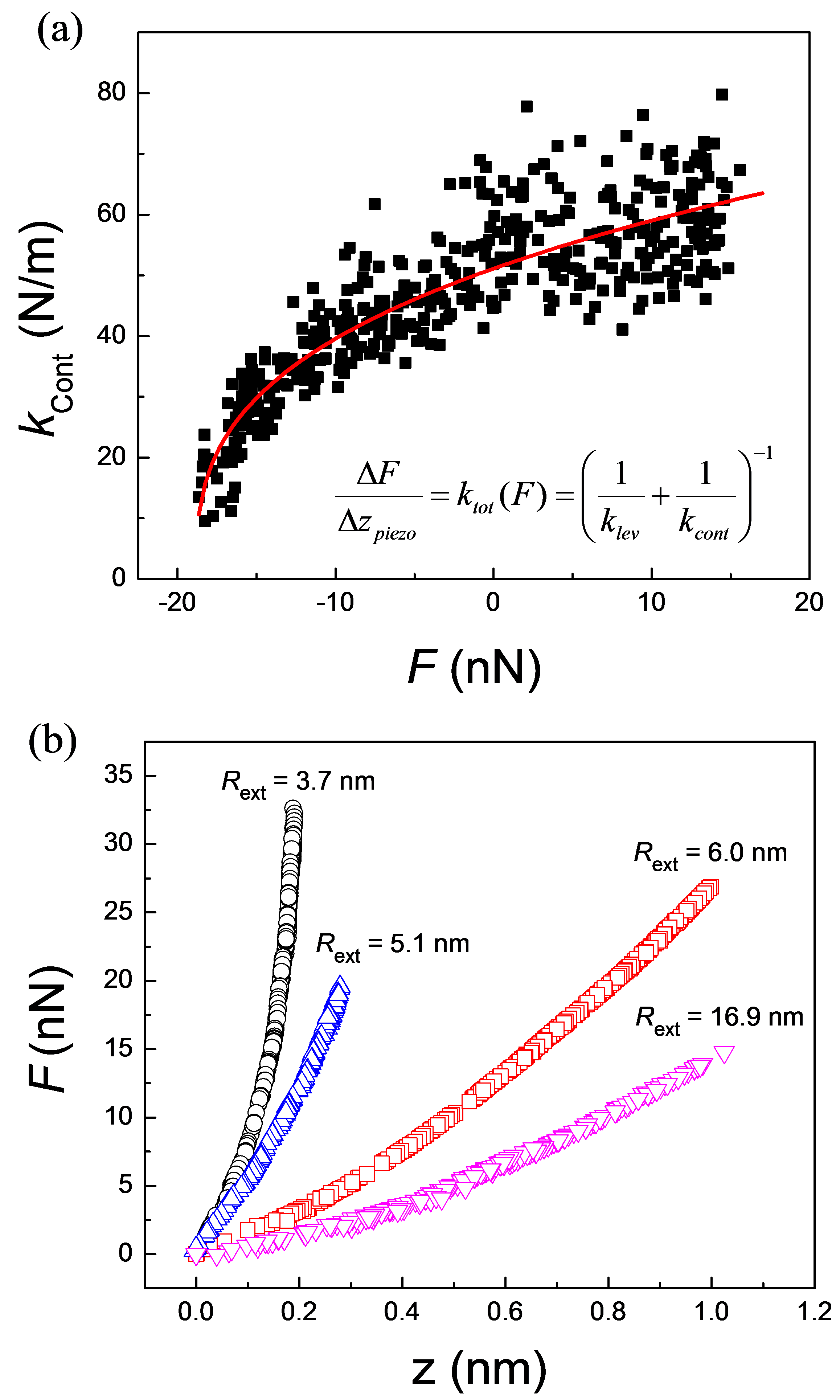}
  \caption{\textit{(a) A typical measured contact stiffness $k_{cont}$ $vs.$ normal force $F$ obtained from a BN-NT of $R_{ext} = $ 13.7 $\pm$ 0.4 nm. The average measured radial elasticity of this BN-NT is 32.0 $\pm$ 3.8 GPa. (b) Normal force $F$ against the indentation depth $z$ on BN-NTs with different $R_{ext}$. The curves are obtained by integrating the experimental data of $1/k_{cont}$ $vs.$ $F$.}}
\end{figure}

To extract the effective radial modulus from MoNI data, we calculate the AFM tip-NT contact area with the Hertz model adding the contribution of the adhesion force $F_{Adh}$, which can be determined directly from data similar to the one illustrated in Fig. 2 (a). For the configuration of a sphere in contact with a cylinder with a normal force $F$ that compresses them together, the Hertz model gives:        

\begin{equation}
\frac{d F}{d z _{piezo}} = k_{cont} = \beta \cdot \left( \frac{4}{3} E^{*} \right)^{2/3} \cdot \left( R \cdot \left( F + F_{Adh} \right) \right)^{1/3}
\end{equation}

where $\frac{1}{R} = \frac{1}{R_{tip}} + \frac{1}{2 R_{ext}}$, $E^{*} = \left( \frac{1-\nu^{2}_{NT}}{E_{Radial}} + \frac{1-\nu^{2}_{tip}}{E_{tip}} \right)^{-1}$, $F_{Adh}$ is the adhesion force between the tip and the nanotube, $\nu_{NT,tip}$ and $E_{Radial,tip}$ are the Poisson's ratio and Young's moduli of respective materials. The parameter $\beta$ takes into account the geometry of the tip-NT contact \cite{16} and more details about the exact calculation of $\beta$ as a function of tip and NT radius are given in Ref. [29]. To better gather an intuition on the meaning of equation (2), it is possible to calculate the elliptical contact area between the tip and the nanotube, and rewrite the relationship between the effective Young modulus and the contact stiffness as $\frac{d F}{d z_{indent}} = k_{cont} = \gamma \sqrt{Area} \cdot E^{*}$, where $Area$ is the contact area, and $\gamma$ is a numerical factor which is function of only $R_{tip}$ and $R_{NT}$ \cite{29}. The radial elastic modulus can thus be calculated using these equations from the experimental measurements of $k_{cont} = d F / d z$. For the calculations, the nominal Poisson's ratio of 0.2 is used for BN-NTs \cite{8,31}. For the silicon AFM tip, we use $\nu_{tip}$ and $E_{tip}$ equal to 0.28 and 130 GPa, respectively. The curves of $k_{cont}$ $vs.$ $F$, as the ones shown in Fig. 2 (a), are then fitted with equation (2) leaving the unknown $E_{Radial}$ as a free fitting parameter. To insure the reliability of our measurements we test this experimental setup on materials with well known Young Modulus, namely fused quartz (72 GPa \cite{32}) and single crystal ZnO [$10\bar{1}0$] (143 $\pm$ 6 GPa \cite{33}). Young's moduli of 71 $\pm$ 18 GPa and 130 $\pm$ 13 GPa are obtained, respectively. 

The measured $E_{Radial}$ for 18 different BN-NTs as a function of $R_{ext}$ is reported in Fig. 3. The radial elastic moduli of CVD MW C-NT obtained by Palaci $et al.$ \cite{16} with the MoNI method and Hertz analysis are also plotted for comparison. The modulus of BN-NT is found to be as large as 192.2 $\pm$ 46.8 GPa for $R_{ext} = $3.7 $\pm$ 0.1 nm, and $L = $ 5.4 $\pm$ 1, and to decrease rapidly as $R_{ext}$ increases until at about $R_{ext} \approx$ 8 nm, and $L \approx$ 12 when $E_{Radial}$ reaches a plateau. For $R_{ext}$ in the range 8 to 36 nm, the average $E_{Radial}$ is found to be 24.1 $\pm$ 11.0 GPa, close to the elastic constant of $h$-BN along its $C_{33}$ axis which is measured to be 27 GPa \cite{26}. This plateau indicates that the radial elasticity eventually approaches a constant value comparable to the modulus of the bulk form of the same material. Indeed, the $E_{Radial}$ plateau value for C-NT is larger and equal to 30 $\pm$ 10 GPa, very close to the Young modulus of HOPG along its $C_{33}$ axis, $E_{HOPG} = $ 30 GPa. These different values can be well understood in terms of the different strength of the non-polar covalent C-C bonds compared to the polar covalent B-N bonds \cite{8}. There is a concern about the deformation of the substrate during the indentation on the nanotubes, especially for stiffer nanotubes. For the smallest and stiffest BN-NT with RNT = 3.7 $\pm$ 0.1 nm, the radial elasticity is measured to be 192.2 $\pm$ 46.8 GPa, with an error of 24.3\% from the fitting procedure. Additional error from cantilever calibration is estimated to be 9\%. If the substrate deforms during indentation, an additional term, $1/k_{NT-substrate}$, needs to be added into equation (1) and the radial elasticity of this BN-NT with the new fitting equation becomes 233.5 $\pm$ 54.2 GPa, showing a 21.5\% increase compared with the previous modulus. However, the differences in radial modulus values considering or neglecting the substrate contribution are within the above mentioned error bar \cite{29}. For the MW NTs whose modulus (neglecting the substrate) is reported in Fig. 3, the average $R_{ext}$/$R_{int}$ is equal to 2.0 $\pm$ 0.48, and 2.2 $\pm$ 0.216, for BN-NTs and C-NTs respectively, as obtained by a TEM statistical analysis of the same samples. Thus it is possible to calculate the number of layers in the BN-NTs as $L = R_{ext}/(2 \cdot d)$, where $d$ is the distance between the walls which is assumed to be 0.34 nm. In Fig. 3, we also plot $E_{Radial}$ $vs.$ $L$ and $vs.$ $R_{int}$ for BN-NTs. The combination of the data reported in Fig. 3 indicates that MW BN-NTs attain the plateau value of $h$-BN along its $C_{33}$ axis at $R_{ext} > $  8 nm or for a number of layers $L > $ 12. On the other hand, C-NTs attain the graphite plateau value for $R_{ext} > $ 4 nm or for $L > $ 5. This different behavior may signify that either BN-NTs have a curvature dependent strain energy which is different than C-NTs and less dramatically dependent on the inverse of the nanotube radius, or the defects play a key role and BN-NTs have a different size dependent defect density than C-NTs. Structural defects such as vacancies in the hexagonal carbon network might reduce the normal rigidity of the carbon sheet, resulting in a reduction of the radial Young's modulus of C-NTs, especially for small nanotubes with only fewer layers. The TEM images indeed confirm that the C-NTs analyzed in Fig. 3 have more structural defects than the BN-NTs reported in the same figure \cite{29}.

\begin{figure}[ht]
  \centering
  \includegraphics[width=0.4\textwidth]{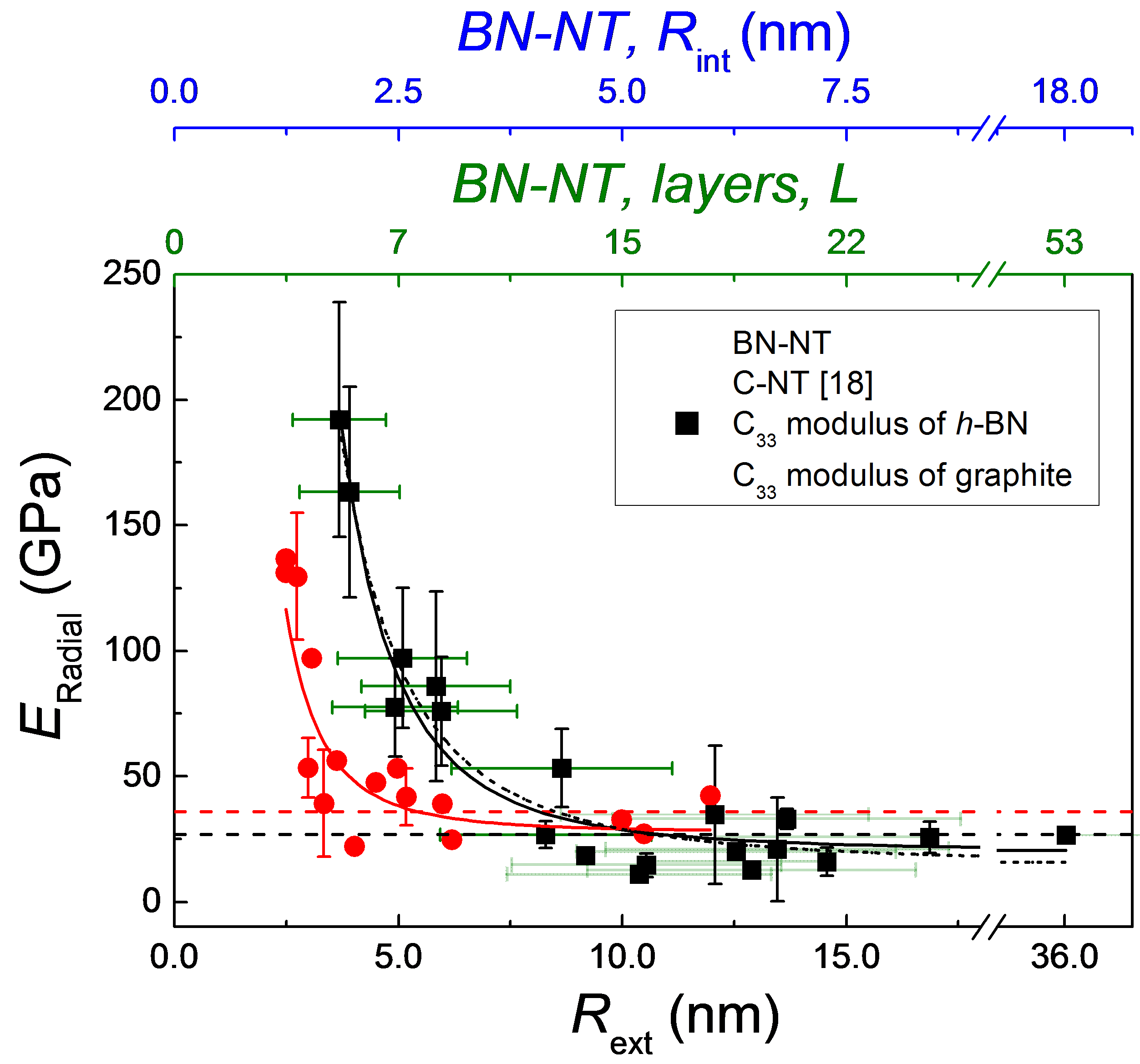}
  \caption{\textit{The measured $E_{Radial}$ of MW BN-NTs investigated in this work (black solid Square).  The vertical (black) and horizontal (green) error bars of $E_{Radial}$ are calculated from averaging multiple MoNI measurements and the uncertainty of the layers in BN-NT, respectively. We use light green color for horizontal error bar for $R_{NT} > $ 9 nm to increase the clarity of the figure. The $E_{NT}$ of MW C-NT (red solid circle) from Ref. [16] is shown for comparison. The solid lines are the best fit to these data with the equation: $E_{Radial} = A/(R_{ext}^{2} (R_{int} - 0.34 nm)^{2} ) + E_{\circ}$. The dotted line is the best fit of the equation: $E_{Radial} = A \sqrt{t} /(R_{ext}^{2} (R_{int} - 0.34)^{2} )) + E_{\circ}$ to the BN-NT data, where $A$ and $E_{\circ}$ are constants.}}
\end{figure}

It is worth comparing the data for BN-NTs presented in this study with the results obtained in a previous study on the radial stiffness of MW BN-NTs \cite{25}. Both studies show that $E_{Radial}$ increases by decreasing the external radius of the nanotubes, however, the most striking observation is that the effective radial moduli of the BN-NTs presented in this letter are much larger than those reported in the previous study \cite{25}. There, $E_{Radial}$ varies between 40 GPa and a few GPa, however these values are found for very small BN-NTs with a number of layers 1 $ < L < $ 4. In particular, they find that the effective radial modulus increases quite dramatically by increasing the number of layers, for a fixed $R_{ext}$. For example, for two BN-NTs with the same $R_{ext} = $ 2.7 nm, the radial modulus is a factor of four larger when the number of layers increases from 3 to 4. In Fig. 3 the smallest BN-NT, with $R_{ext} = $3.7 $\pm$ 0.1 nm, and $L = $5.4 $\pm$ 1, presents a radial modulus of 192.2 $\pm$ 46.8 GPa, more than twenty times larger than the modulus measured in the previous study for a nanotube with similar external radius but a smaller number of layers. This comparison clearly shows that the increased radial stiffness is related to a larger number of layers in the BN-NTs studied here, furthermore, we remark that larger L values for a fixed $R_{ext}$ mean a decrease of the internal radius, which contributes to increase the radial stiffness of the nanotube. However, although a larger thickness increases the radial stiffness, the data reported in Fig. 3, where the thickness of the BN-NT is proportional to $R_{ext}$, in fact $t = 0.34 \cdot L = 0.5 \cdot R_{ext}$, show that the decrease of the modulus due to an increase of $R_{ext}$ and $R_{int}$ is predominant over the increase due to the thickness. Such a strong dependence on the curvature can be ascribed to the combination of the stress along the tube radius (in the direction of the wall thickness) and the stress parallel to the tube surface during the local radial indentation. Since the in plane elastic constants are about 2 orders of magnitude larger than the 33 constants in BN and graphite, it appears clear that the same indentation will produce a larger strain on the in plane spring constants in a smaller NT than in a larger one. In an effort to better understand the data of Fig. 3, we fit them with the function $E_{Radial} = A/(R_{ext}^{2} (R_{int} - 0.34 nm)^{2}) + E_{\circ}$, where $A$ and $E_{\circ}$ are two constant values. This function accounts for the increase of the tube's strain energy with $1/R_{ext}^{2}$, for the dramatic increase in $E_{Radial}$ when $R_{int}$ approaches the interlayer distance between graphene layers, and for the approaching to the modulus of the bulk material for large and thick tubes. Similarly, a theoretical study has shown that for MW C-NTs, if the internal radius is fixed to 0.34 nm, the effective radial modulus decreases nonlinearly with increasing external radius, reaching a plateau (47 GPa) at $Rext > $ 10 nm \cite{34}. The fit to the data in Fig. 3 gives $E_{\circ} = $20.3 $\pm$ 3.4 GPa for the BN-NTs and $E_{\circ} = $27.9 $\pm$ 6.0 GPa for C-NTs with similar morphology. These values are very close to the elastic constants along the $C_{33}$ axis of the corresponding bulk materials, i.e., $h$-BN \cite{26} and HOPG. We note that since $t$ is proportional to $R_{ext}$ and $R_{int}$ in our BN-NTs, we cannot address precisely the role of the thickness here. An alternative fitting function, also shown in Fig.3 as dotted line, is given by $E_{Radial} = A \sqrt{t} / (R_{ext}^{2} (R_{int} - 0.34)^{2} ) + E_{\circ}$.

In summary, when the indentation is small compared to the external radius of the nanotube, and for thick shell tubes, the localized radial compression gives rise to minimal ovalization, whereas it creates a dimple in the tube layers and the elasticity is strongly dependent on the layer-layer interaction and layer curvature, such that for a large number of layers and a large radius of curvature the radial elastic modulus collapses to the transverse elastic modulus of the corresponding two dimensional material, e.g. graphite or BN films. We find that the effective radial modulus of MW BN-NTs with thick walls is mainly controlled by the internal and external radii and it decreases rapidly as the nanotube's external radius increases. The radial modulus decreases from about 192 GPa when $R_{ext}$ is 3.7 $\pm$ 0.1 nm and has 5.4 $\pm$ 1 layers, to an asymptotic value of 24 $\pm$ 11 GPa for $R_{ext}$ larger than 8 nm, approaching the modulus of $h$-BN along its $C_{33}$ axis. Compared to MW C-NTs, the asymptotic value of $E_{Radial}$ of MW BN-NTs is reduced by approximately 20\%, in agreement with graphite's transverse elastic constants compared to BN. These nanotubes have potential applications in nanoscale devices such as actuators and sensors as well as reinforcement for both ceramic and polymer composite materials. Compared to C-NTs, the radially soft but the axially robust MW BN-NTs might be better candidates for composites reinforcement when high temperature stability and electrical insulating capability are required \cite{35}. In addition, it is known that BN-NTs interact strongly with polymers due to the electrical polarization in BN-NTs induced by broken symmetry \cite{4}. Their radial softness might result in better conformation and dispersion of BN-NTs in the polymer matrix. Moreover, the flexible MW BN-NTs are advantageous in reinforcing bio-degradable polymers for orthopedic implant applications because their softness will not adversely affect the ductility of the scaffolds \cite{36,37}. 

H.-C.C., S.K., and E.R. acknowledge the financial support of the Office of Basic Energy Sciences of the US Department of Energy DOE (DE-FG02-06ER46293). E.R. acknowledges the National Science Foundation NSF (DMR-0820382 and CMMI-1100290) for partial support.  The authors also thank Dr. Antonio Pantano and Dr. Ken Gall for useful discussions.


\end{document}